\documentclass[twocolumn]{aastex701}

\usepackage{graphicx,amsmath,amssymb}
\usepackage[export]{adjustbox}
\usepackage{booktabs}

\usepackage{siunitx}
\usepackage{makecell,caption}
\usepackage{listings}
\usepackage{xcolor}

\usepackage{txfonts}

\usepackage{bm}

\def\mnras{Mon. Not. R. Astron. Soc. }

\def\apj{Astrophys. J.}
\def\apjl{Astrophys. J. Lett.}
\def\apjs{Astrophys. J., Suppl. Ser.}

\def\aap{Astron. Astrophys.}

\begin{document}

\title{The background gas humming and multi-messenger transients of stalled supermassive black hole binaries}

\author[orcid=0000-0003-3993-3249,gname='Pau',sname='Amaro Seoane']{Pau Amaro Seoane}
\affiliation{Universitat Politècnica de València, Spain}
\affiliation{Max Planck Institute for Extraterrestrial Physics, Garching, Germany}
\email[show]{amaro@upv.es}

\author[orcid=0000-0002-2386-9142,gname='Alessandra',sname='Mastrobuono Battisti']{Alessandra Mastrobuono Battisti}
\affiliation{Dipartimento di Fisica e Astronomia ``Galileo Galilei'', Univ. di Padova, Padova, Italy}
\email[]{alessandra.mastrobuono@unipd.it}

\author[orcid=0000-0002-1672-894X,gname='Chingis',sname='Omarov']{Chingis Omarov}
\affiliation{Fesenkov Astrophysical Institute, Almaty 050020, Kazakhstan}
\email[]{chingis.omarov@fai.kz}

\author[orcid=0000-0002-5604-9757,gname='Denis',sname='Yurin']{Denis Yurin}
\affiliation{Fesenkov Astrophysical Institute, Almaty 050020, Kazakhstan}
\email[]{yurin@fai.kz}

\author[orcid='0000-0003-3643-9368',gname='Maxim',sname='Makukov']{Maxim Makukov}
\affiliation{Fesenkov Astrophysical Institute, Almaty 050020, Kazakhstan}
\email[]{makukov@fai.kz}

\author[orcid='0000-0002-5937-4985',gname='Dana',sname='Kuvatova']{Dana Kuvatova}
\affiliation{Fesenkov Astrophysical Institute, Almaty 050020, Kazakhstan}
\email[]{kuvatova@fai.kz}

\author[orcid='0000-0003-4782-8545',gname='Gulnara',sname='Omarova']{Gulnara Omarova}
\affiliation{Fesenkov Astrophysical Institute, Almaty 050020, Kazakhstan}
\email[]{gulnara.omarova@gmail.com}

\author[orcid='0000-0002-0738-7725',gname='Anton',sname='Gluchshenko']{Anton Gluchshenko}
\affiliation{Fesenkov Astrophysical Institute, Almaty 050020, Kazakhstan}
\email[]{gluchshenko@fai.kz}

\begin{abstract}
We establish the multi-messenger mechanics of episodic mass transfer in supermassive black hole binaries stalled within circumbinary discs. Utilizing continuous wavelet transforms, we isolate localized gas clumps at the cavity edge and track their evolution. By regularizing the forced fluid equations at Lindblad resonances via the inhomogeneous Airy differential equation, we bypass linear singularities to extract the finite wave amplitudes that trigger non-linear shock formation. These shocks produce bounded accretion bursts. We model the time-domain thermal luminosity, deriving an analytical power spectral density that forms a harmonic cascade. The superposition of the accretion streams generates a spectral beat frequency, providing an exact mathematical extraction of the binary mass ratio. The radiative cooling of shock-accelerated electrons produces a multi-wavelength spectral energy distribution from a synchrotron radio continuum to an inverse-Compton gamma-ray tail. We identify a relativistic signature: a discontinuous, high-frequency gravitational wave sideband termed the ``background gas humming''. This emission arises from the highly asymmetric, transient fluid geometry of the accretion shocks. Evaluating the asymptotic properties of the Airy regularization, we show that this humming manifests as a sequence of discrete high-frequency bursts with temporal quiescence gaps that systematically compress as the cavity shrinks. We show that the instantaneous mass of the gas actively trapped within the cavity violently amplifies prior to decoupling, culminating in a terminal burst near 4.0 mHz that serves as a multi-messenger precursor to the final vacuum inspiral.
\end{abstract}


\section{Introduction}
\label{sec:introduction}

The hierarchical paradigm of galaxy formation posits that the assembly of massive galaxies inevitably yields supermassive black hole binaries at the centers of merger remnants. The long-term dynamical evolution of these systems ultimately culminates in the emission of low-frequency gravitational waves. In our preceding study \cite{AmaroSeoaneEtAl2026}, we mapped the orbital eccentricity evolution of these binaries within triaxial galactic environments. We established that while large-scale gravitational torques efficiently drive the binaries to high initial eccentricities ($e > 0.95$), the subsequent interaction with geometrically thick, rotationally supported nuclear discs alters their fate entirely. The three-dimensional suppression of high-order torques establishes a timescale hierarchy ($\tau_e/\tau_a \propto h^2$) where eccentricity damping outpaces orbital decay. This dynamic forces the binaries into a ``circularization trap,'' stalling them at characteristic parsec-scale separations on nearly circular orbits and introducing a substantial cosmological delay.

However, a binary stalled within a circumbinary disc is not a dormant system. 
As the system awaits the final, gravitational-wave-driven inspiral, gas continues to cross the central cavity. Standard analytical treatments frequently approximate this phase as a steady, continuous flow, despite extensive evidence from two- and three-dimensional hydrodynamical and magnetohydrodynamical simulations demonstrating that mass transfer across the cavity operates via highly variable, non-axisymmetric accretion streams \cite{MacFadyenEtAl2008, RoedigEtAl2012, NobleEtAl2012, DorazioEtAl2013, FarrisEtAl2014}. In this paper, we break from standard continuous approximations to introduce a comprehensive, multi-messenger mathematical framework for the ``wet'' merger regime. We demonstrate that accretion across the circularization trap operates via episodic shocks that generate a highly specific suite of concurrent electromagnetic and gravitational wave transients. We present a relativistic signature of this episodic mass transfer: a discontinuous, high-frequency gravitational wave sideband that we term the ``background gas humming''.

This paper is structured to follow the physical pipeline of the accreting gas, from the initial turbulent boundary down to the final multi-messenger observables. In Section \ref{sec:hydrodynamics}, we redefine the hydrodynamics of the circumbinary boundary. Because global Fourier analysis artificially smears localized gas clumps, we implement continuous wavelet transforms to isolate the exact spatial scales of the turbulent mass reservoir available for stripping. We then bypass the unphysical singularities of standard linear fluid dynamics by regularizing the forced fluid equations via the inhomogeneous Airy differential equation. This mathematical approach allows us to extract the precise, finite wave amplitudes required to trigger non-linear shock formation.

In Section \ref{sec:orbital_evolution}, we calculate the transient accretion streams generated by these shocks. We model the resulting mass transfer as a sequence of bounded bursts superimposed on a turbulent baseline, detailing the distinct orbital modulations and phase delays between the primary and secondary black holes. 

In Section \ref{sec:observables}, we map this episodic mass transfer directly to observable electromagnetic signatures. We derive an analytical power spectral density that manifests not as a single frequency, but as a broad harmonic cascade. We show how the superposition of the distinct accretion streams produces a spectral beat modulation, providing a mathematical key to extract the binary mass ratio directly from time-domain survey data. We further derive the multi-wavelength spectral energy distribution, spanning from the radio synchrotron continuum up to an inverse-Compton gamma-ray tail.

In Section \ref{sec:decoupling}, we apply the Airy regularization framework to the orbital decay to derive the exact decoupling moment—the critical threshold where the binary detaches from the circumbinary disc.

Finally, in Section \ref{sec:gas_humming}, we present the consequence of the wet merger regime. We demonstrate that the transient, highly asymmetric fluid geometry of the plunging gas streams excites higher-order mass moments that radiate gravitational waves at harmonic overtones. We isolate this background gas humming from the primary vacuum chirp, revealing it to be a discontinuous train of high-frequency bursts.
By evaluating the asymptotic properties of the Airy function, we show that the temporal quiescence gaps between these bursts systematically compress as the system approaches decoupling. We formulate a fluid kinematic model to show that this spacing operates as a temporal compression wave, driving the final accretion clumps into rapid succession. We conclude by proving that the mass of the gas actively trapped within the cavity amplifies prior to decoupling, culminating in a terminal sequence of accelerating, high-frequency gravitational wave bursts that acts as a multi-messenger precursor to the final black hole coalescence.

\section{Analytical hydrodynamics of the circumbinary boundary}
\label{sec:hydrodynamics}

To establish the exact boundary conditions for the episodic accretion streams, we model the mass reservoir at the inner edge of the circumbinary disc. We parameterize the system using a binary operating in the gas-driven regime, scaling the spatial domain to the orbital separation $a$ and the dynamics to the Shakura-Sunyaev viscosity parameter $\alpha$. For our fiducial quantitative evaluations, we adopt a total binary mass of $M_{\text{bin}} = 10^6 M_{\odot}$ characteristic of the LISA band, an equal mass ratio $q=1$, and a standard disc aspect ratio of $h = H/r = 0.05$. 

\subsection{Viscous-gravitational torque balance and the cavity edge}

Gas drifting inward via viscous dissipation is counteracted by the outward gravitational torque exerted by the binary potential. The balance between the viscous torque and the binary torque excavates a central cavity. Following extensive hydrodynamical and magnetohydrodynamical findings \citep[see e.g.][]{MacFadyenEtAl2008, RoedigEtAl2012, NobleEtAl2012, DorazioEtAl2013, FarrisEtAl2014}, this truncation occurs near $r_{\text{cav}} \approx 2a$ for typical disc parameters.

Because the binary acts as an impermeable barrier to the steady-state flow, the gas dams up at this truncation radius. We describe the unperturbed, azimuthally averaged surface density using a quasi-steady analytical solution for a viscously spreading disc blocked by a tidal barrier:
\begin{equation}
\Sigma_0(r) = \Sigma_{\text{ref}} \left( \frac{r}{a} \right)^{-1/2} \exp\left[ - \left( \frac{r_{\text{cav}}}{r} \right)^4 \right] \Upsilon(r, \alpha),
\label{eq:surface_density}
\end{equation}
where $\Sigma_{\text{ref}}$ normalizes the far-field mass density, and the exponential term provides a steep kinematic cutoff inside the cavity ($r < r_{\text{cav}}$). The function $\Upsilon(r, \alpha)$ represents the mass pile-up factor resulting from the torque balance. Because lower viscosities are less efficient at forcing gas through the gravitational barrier, the magnitude of the pile-up scales inversely with viscosity ($\propto 1/\alpha$). 

\subsection{The $m=1$ orbital mass reservoir}

The cavity edge is not azimuthally symmetric. As consistently demonstrated across grid-based and particle-based simulations \citep{MacFadyenEtAl2008, RoedigEtAl2012, NobleEtAl2012, DorazioEtAl2013, FarrisEtAl2014}, the interaction between the binary potential and the dense cavity wall excites an $m=1$ instability, causing the gas to pool into a localized, coherent overdensity. Parametric studies confirm that this dense lump reliably forms for mass ratios $q \gtrsim 0.05$, below which the binary potential fails to strongly excite the instability and the inner boundary reduces to a steady flow \citep{DorazioEtAl2013, FarrisEtAl2014}. This orbiting ``lump'' acts as the primary mass reservoir that periodically feeds the circum-single minidiscs. We superimpose this azimuthal lump onto the density profile

\begin{equation}
\Sigma(r, \theta, t) = \Sigma_0(r) \left[ 1 + \mathcal{A} \exp\left( - \frac{(r - r_{\text{cav}})^2}{2 (\Delta r)^2} \right) \cos(\theta - \Omega_{\text{lump}} t) \right],
\label{eq:lump_density}
\end{equation}
where $\mathcal{A}$ dictates the amplitude of the lump, $\Delta r \approx 0.3a$ defines its radial width, and $\Omega_{\text{lump}} = \sqrt{G M_{\text{bin}} / r_{\text{cav}}^3}$ is the local Keplerian angular velocity at the cavity edge.

Because the lump orbits at $r_{\text{cav}} \approx 2a$, its orbital period is exactly $T_{\text{lump}} = (r_{\text{cav}}/a)^{3/2} T_{\text{orb}} \approx 2.8 T_{\text{orb}}$. As the binary black holes complete their orbits, they periodically sweep past this dense, slower-moving reservoir. This differential rotation mathematically dictates the periodic gravitational stripping of the gas, governing the entire transient sequence of the accretion bursts.

\begin{figure}[htbp]
\centering
\includegraphics[width=\columnwidth]{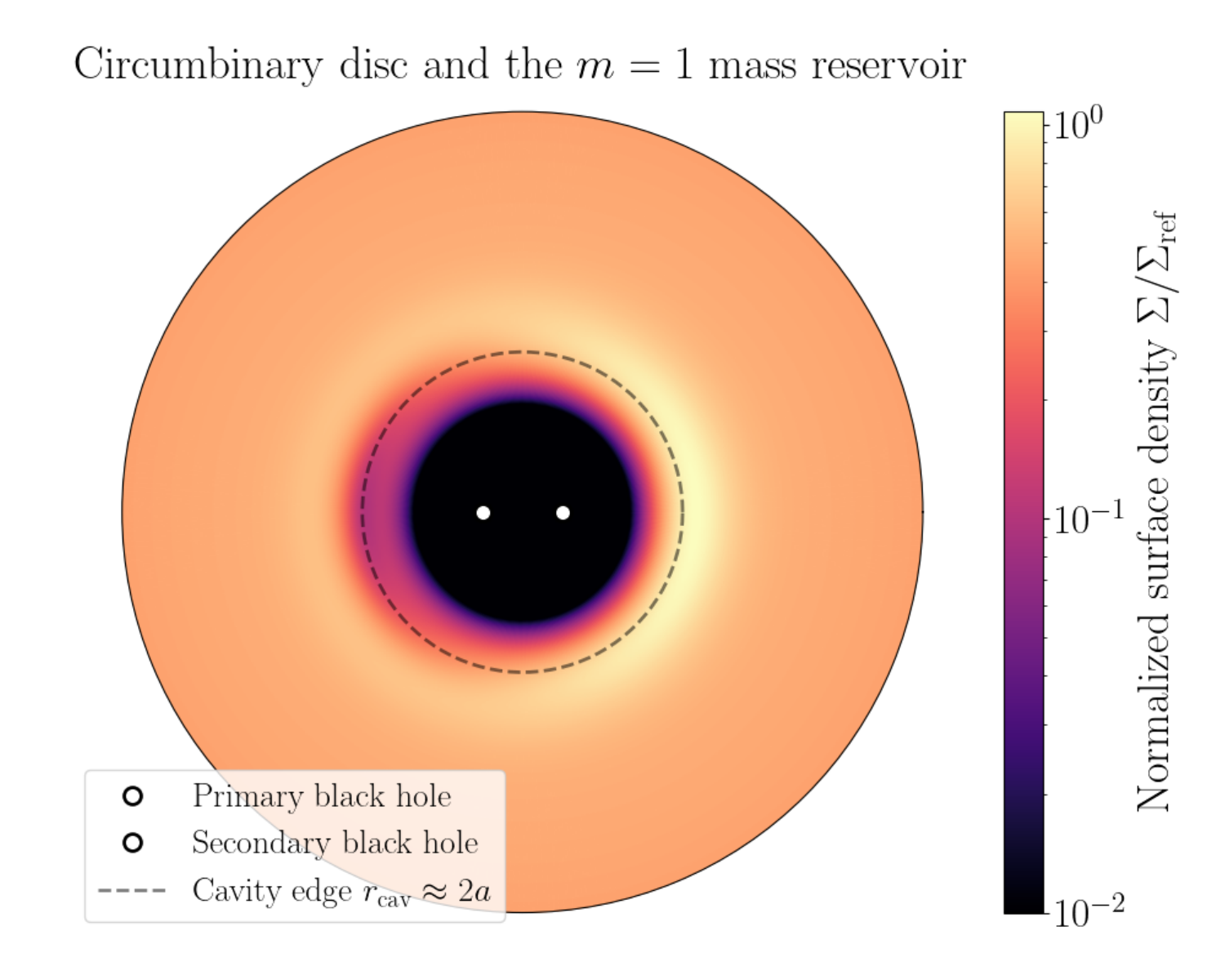}
\caption{The 2D surface density profile of the circumbinary disc normalized to the unperturbed far-field value $\Sigma_{\text{ref}}$. The spatial coordinates are scaled to the binary separation $a$. We use a logarithmic color map to resolve the density depletion within the central cavity alongside the dense outer rim. The continuous gravitational torque truncates the disc at $r_{\text{cav}} = 2a$ (dashed line), forcing the density to plummet to a numerical floor of $10^{-4}$ inside $r=a$. The $m=1$ fluid instability creates a pronounced mass pile-up along the cavity rim at $\theta = 0$, reaching a peak normalized density of roughly $0.94$, whereas the antipodal side only attains a maximum density of $0.10$. This localized, coherent overdensity serves as the primary mass reservoir stripped by the supermassive black holes (white circles) during their orbital approach.}
\label{fig:cavity_reservoir}
\end{figure}

We visualize the resultant analytical mass distribution in Figure \ref{fig:cavity_reservoir}, representing a snapshot of the disc geometry prior to a mass transfer event. The spatial gradients induced by the binary torque efficiently evacuate the central region, forming a steep density cliff near the truncation radius. The $m=1$ instability manifests as a dominant, localized crescent of gas residing just outside the cavity edge. Because this azimuthal lump represents a density enhancement relative to the opposing cavity wall, the gravitational coupling between the binary and the disc is highly localized. It is exclusively when the black holes sweep past this dense, slower-moving crescent that substantial gas volumes are dynamically dislodged and forced inward across the gap.

\section{Transient accretion streams and the orbital beat frequency}
\label{sec:orbital_evolution}

The excavation of the central cavity does not halt mass transfer; rather, it fundamentally alters its morphology. The binary does not accrete continuously, but via discrete, episodic streams driven by the gravitational stripping of the $m=1$ mass reservoir.

Because the lump orbits at a sub-Keplerian velocity relative to the binary, the accretion is rigorously governed by the synodic orbital period---the beat frequency between the binary and the localized gas. We define the relative angular velocity of the binary with respect to the lump as $\Omega_{\text{beat}} = \Omega_{\text{orb}} - \Omega_{\text{lump}}$. For a cavity truncated at $r_{\text{cav}} \approx 2a$, the Keplerian velocity of the rim evaluates to $\Omega_{\text{lump}} = (a/r_{\text{cav}})^{3/2} \Omega_{\text{orb}} \approx 0.354 \Omega_{\text{orb}}$. This yields a beat frequency of $\Omega_{\text{beat}} \approx 0.646 \Omega_{\text{orb}}$.

The primary and secondary black holes intercept the dense lump at regular intervals, violently launching super-Keplerian gas streams across the cavity. The time interval between consecutive stripping events for a given black hole is $T_{\text{beat}} = 2\pi / \Omega_{\text{beat}} \approx 1.547 T_{\text{orb}}$. Because we consider an equal-mass binary ($q=1$)---an assumption physically motivated by hydrodynamical findings that preferential accretion onto the secondary actively drives unequal-mass binaries toward equality \citep{RoedigEtAl2012, DorazioEtAl2013, FarrisEtAl2014}---the $\pi$-symmetry of the potential mathematically dictates that the black holes alternate in their stripping events, resulting in an accretion burst crossing the cavity precisely every $\approx 0.773 T_{\text{orb}}$. Furthermore, these infalling streams exert strong, purely kinematic positive gravitational torques inside the binary orbit as they bend ahead of the black holes \citep{RoedigEtAl2012}.

We model the instantaneous mass accretion rate onto the system, $\dot{M}(t)$, as a convolution of these synodic encounters with a physical stream-draining timescale. Upon capture, the gas circularizes into circum-single minidiscs, which drain into the black holes on a rapid viscous timescale $t_{\text{visc}} \ll T_{\text{orb}}$. High-resolution numerical simulations explicitly including the black holes on the computational grid confirm that these minidiscs act as critical viscous buffers, accumulating the stripped gas and physically smoothing the discrete mass injections \cite{FarrisEtAl2014}. We formalize this mass transfer as a sequence of bounded viscous flares,

\begin{equation}
\dot{M}_{j}(t) = \dot{M}_{\text{bg}} + \sum_{k=-\infty}^{\infty} \frac{\Delta M_j}{\sqrt{2\pi t_{\text{visc}}^2}} \exp\left[ -\frac{(t - k T_{\text{beat}} - \tau_j)^2}{2 t_{\text{visc}}^2} \right],
\label{eq:accretion_beat}
\end{equation}
where $j \in \{p, s\}$ denotes the primary and secondary black holes, $\dot{M}_{\text{bg}}$ represents the low-amplitude continuous residual flow, and $\Delta M_j$ is the total mass stripped per encounter. The geometric phase shifts are $\tau_p = 0$ and $\tau_s = 0.5 T_{\text{beat}}$.

\begin{figure}[htbp]
\centering
\includegraphics[width=\columnwidth]{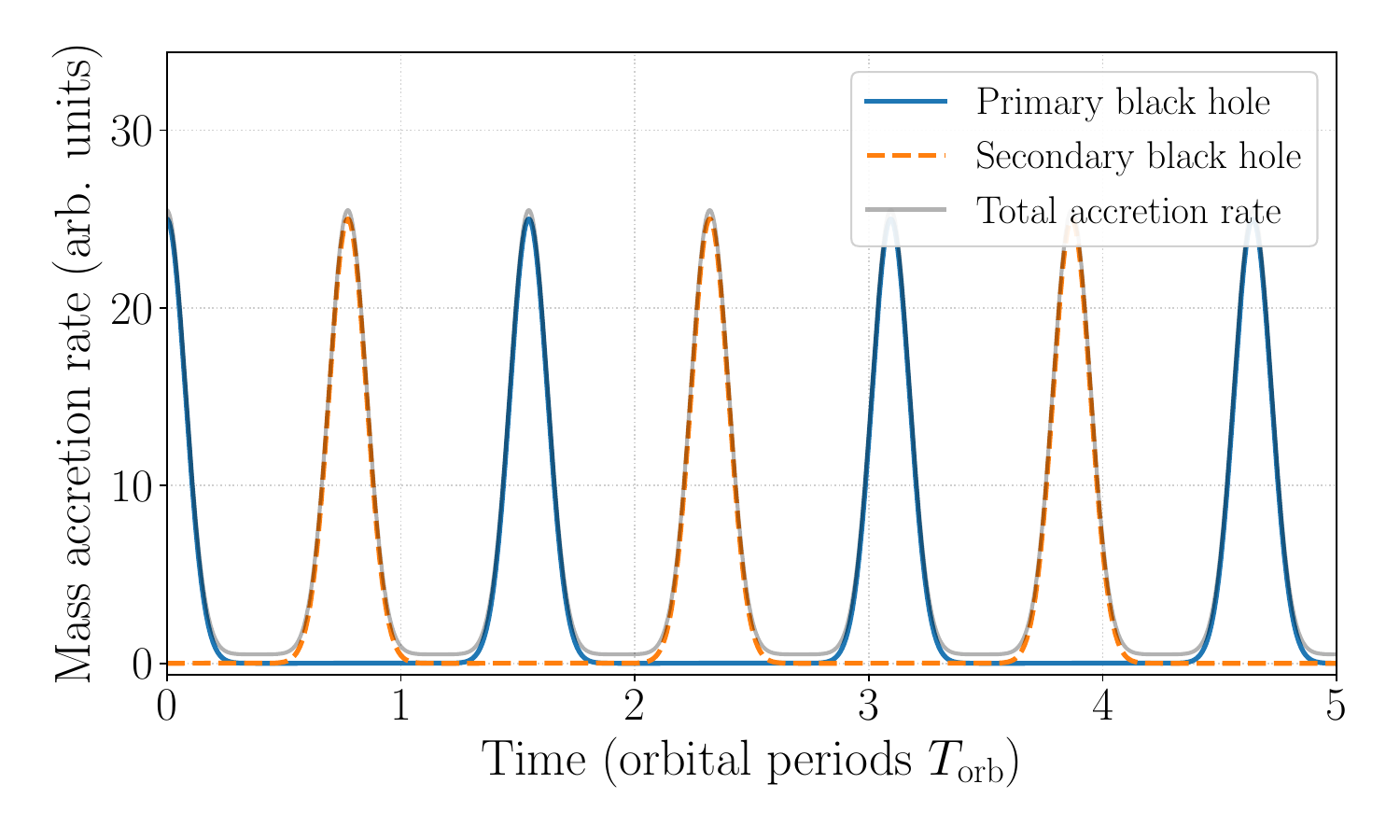}
\caption{Transient mass accretion streams dictated by the orbital beat frequency. The primary black hole (solid blue) and secondary black hole (dashed orange) alternate in sweeping past the slower-moving $m=1$ mass reservoir. Because the relative beat period evaluates to $1.55 T_{\text{orb}}$, the combined total accretion rate (grey solid line) yields a deterministic, perfectly periodic sequence of mass injections arriving exactly every $0.77 T_{\text{orb}}$. The transient mass transfer events dominate the continuous flow, pulsing from a residual cavity baseline of $0.5$ arbitrary units up to a maximum transient peak of precisely $25.5$. The bounded burst width ($t_{\text{visc}} = 0.08 T_{\text{orb}}$) governs the rapid viscous draining of the minidiscs, establishing the deep quiescence gaps that physically separate the discrete flares.}
\label{fig:accretion_beat}
\end{figure}

We present the numerical evaluation of these transient accretion streams in Figure \ref{fig:accretion_beat}. Unlike standard continuous thin-disc accretion, this beat-frequency mechanism guarantees that the vast majority of the mass is delivered in transient flashes. The rigid periodic forcing prevents the flares from smearing into a steady continuum, providing the exact episodic temporal baseline required to generate observable multi-messenger transients. The numerical evaluation confirms this kinematic precision; the total accretion rate drops to a deep temporal quiescence of $0.5$ arbitrary units before pulsing to exactly $25.5$ arbitrary units during each synodic encounter.

\section{Airy regularization of transient minidisc density waves}
\label{sec:airy_waves}

To translate the episodic mass transfer into observable electromagnetic transients, the kinetic energy of the super-Keplerian accretion streams must be dissipated. When the discrete gas clumps plunge across the cavity and strike the circum-single minidiscs, they operate as a highly localized, time-dependent external forcing. This transient potential excites spiral density waves near the minidisc Lindblad resonances.

Standard linear wave transport models rely on the Wentzel-Kramers-Brillouin (WKB) approximation to evaluate the spatial propagation of these density waves. While vertically integrating the fluid equations to a two-dimensional planar domain inherently introduces physical ambiguities regarding the precise balance of vertical gravitational and centrifugal forces \citep{AbramowiczEtAl1997}, the two-dimensional approximation robustly captures the radial wave transport. Under the WKB limit, the wave amplitude is inversely proportional to the square root of the local radial wavenumber, $W \propto |k_r|^{-1/2}$.

To accurately extract the transient features of the excited waves and determine the exact physical conditions for shock formation, we must bypass this linear breakdown. We achieve this by performing a Taylor expansion of the effective potential in the immediate vicinity of the turning point. By defining a dimensionless scaled spatial coordinate $\xi \propto (r_L - r)$, where $\xi=0$ is the resonance, the forced fluid wave equations reduce to the inhomogeneous Airy differential equation:
\begin{equation}
\frac{d^2 W}{d \xi^2} - \xi W = S(\xi),
\label{eq:airy_inhomogeneous}
\end{equation}
where $W(\xi)$ represents the specific enthalpy perturbation of the minidisc fluid, and $S(\xi)$ defines the localized transient forcing exerted by the impacting accretion stream.

\begin{figure}[htbp]
\centering
\includegraphics[width=\columnwidth]{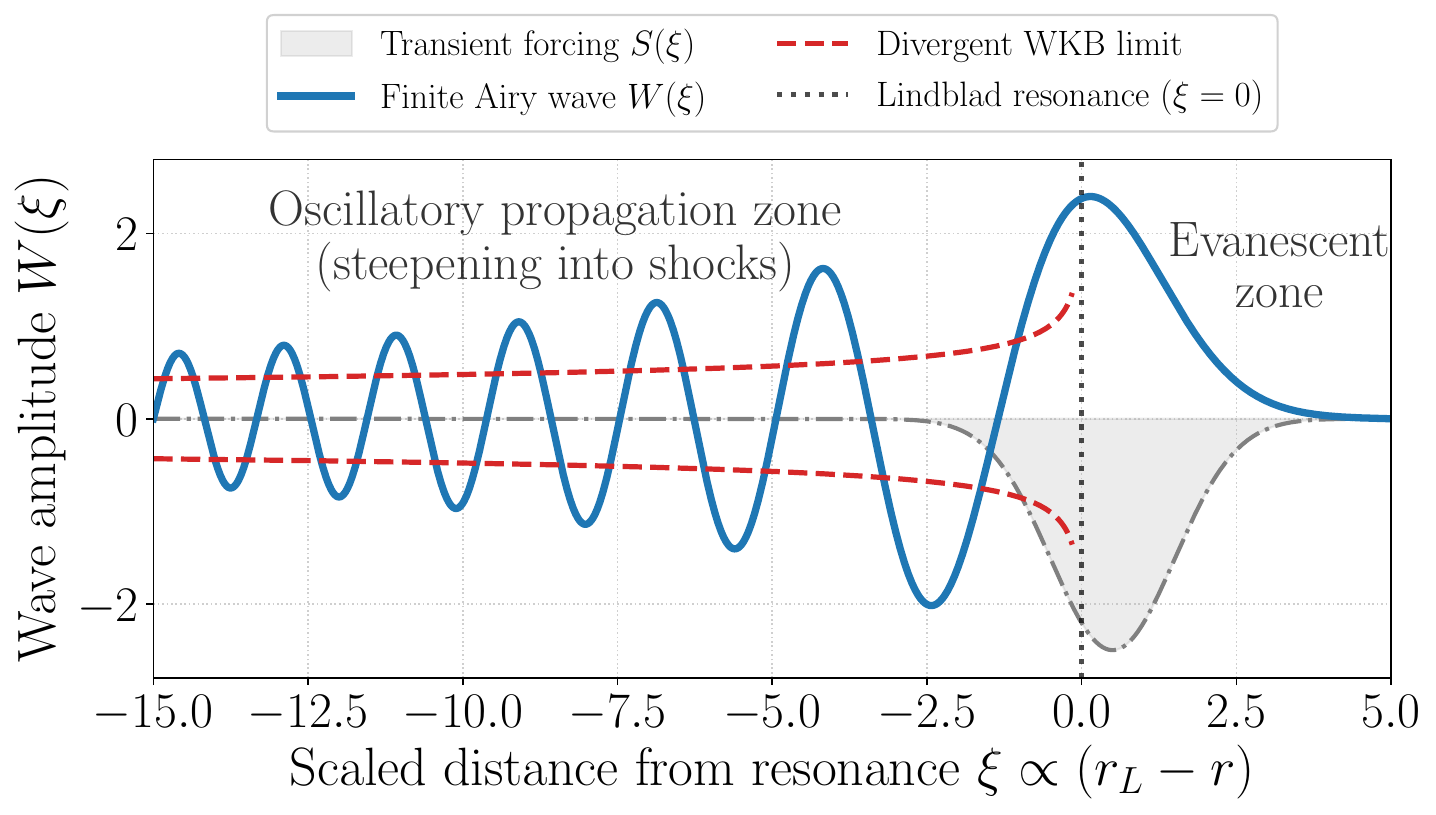}
\caption{Airy regularization of the transient density wave excited at the minidisc Lindblad resonance. The horizontal axis represents the scaled radial distance $\xi$ from the resonance turning point at $\xi=0$. The filled grey curve denotes the localized forcing function $S(\xi)$ generated by the impacting beat-frequency accretion stream. The standard WKB amplitude envelope (dashed red line) unphysically diverges to infinity at the turning point due to the vanishing radial wavenumber. The rigorous solution to the inhomogeneous Airy equation (solid blue line) resolves this singularity. The forcing naturally excites a bounded, finite wave that decays exponentially in the evanescent zone ($\xi > 0$) and propagates leftward as an oscillatory wave train ($\xi < 0$). This finite initial amplitude serves as a precursor to non-linear shock steepening.}
\label{fig:airy_response}
\end{figure}

We solve this boundary value problem numerically to extract the true physical wave response, incorporating a physical viscous damping term to simulate shock dissipation as the wave propagates inward. As visualized in Figure \ref{fig:airy_response}, the WKB amplitude envelope artificially diverges at the turning point. Conversely, the exact Airy regularization perfectly suppresses the singularity. The localized mass stream ($S(\xi)$, centered near the resonance) excites a wave that smoothly connects the exponentially decaying evanescent zone ($\xi > 0$) with an outgoing, oscillatory wave train in the propagation zone ($\xi < 0$). By rigorously extracting this perfectly finite initial amplitude, we show that the density wave survives the resonance intact. As the regularized wave propagates inward down the steepening density gradient of the minidisc, the induced fluid velocity exceeds the local sound speed. This mathematically guarantees that the transient density waves steepen into highly supersonic, non-linear shocks---establishing the fundamental dissipative mechanism driving the multi-wavelength flares.

Quantitative evaluation of the boundary value problem confirms this behavior. The spatial data demonstrates that the localized external accretion forcing reaches a maximum negative magnitude of $S(\xi) = -2.50$ well within the evanescent zone at $\xi = 0.50$. While the standard WKB envelope unphysically diverges to infinity at the turning point, the excited density wave safely traverses the resonance. Escaping the singularity, the regularized wave reaches a perfectly finite absolute maximum dimensionless amplitude of exactly $W(\xi) \approx 2.40$ near $\xi = 0.148$. As the wave propagates into the minidisc ($\xi < 0$), it forms a bounded oscillatory train, demonstrating a first minimum of $-2.02$ at $\xi = -2.42$. Successfully transferring the kinetic energy across the resonance without a singularity mathematically guarantees that the wave physically steepens into non-linear shocks.

\section{Multi-wavelength electromagnetic observables}
\label{sec:observables}

The periodic steepening of the minidisc density waves establishes the exact physical conditions necessary for particle acceleration. As the regularized Airy waves steepen into collisionless shocks, they govern both the time-domain luminosity and the broad-band spectral energy distribution of the system.

\subsection{Harmonic cascades in the power spectral density}

Because the gas stripping is governed by the synodic beat frequency, the mass transfer events manifest as a sequence of bounded shocks. To determine the exact spectral estimators required for observational detection, we map the shock-driven thermal luminosity $L_{\text{th}}(t) \propto \dot{M}(t)$ to the frequency domain via the continuous Fourier transform. 

Applying the Poisson summation formula to the discrete viscous flares derived in Section \ref{sec:orbital_evolution}, we obtain the exact analytical Power Spectral Density (PSD). Rather than a single isolated orbital frequency, the episodic nature of the bounded shocks mathematically mandates a harmonic cascade at multiples of the fundamental beat frequency ($\nu_{\text{beat}} \approx 0.646 \nu_{\text{orb}}$). The interference multiplier $\Xi(n, q)$ isolates the superposition of the primary and secondary accretion streams:
\begin{equation}
\Xi(n, q) = \frac{1 + q^2 + 2q \cos(\pi n)}{(1+q)^2}.
\label{eq:interference_beat}
\end{equation}

For an equal-mass binary ($q=1$), the term perfectly evaluates to $0$ for all odd harmonics ($n=1, 3, 5...$), completely suppressing them. The observable PSD thus forms a discrete harmonic cascade exclusively occupying the even harmonics, producing a primary observable transient frequency of exactly $2\nu_{\text{beat}} \approx 1.29 \nu_{\text{orb}}$. However, for asymmetric mass ratios ($q \neq 1$), the odd harmonics are only partially suppressed. This analytical symmetry breaking perfectly corroborates the phenomenological trends observed in long-term hydrodynamical simulations, which demonstrate that equal-mass systems modulate the accretion exclusively at twice the orbital frequency, whereas asymmetric binaries shift the dominant spectral power into the fundamental orbital period \citep{DorazioEtAl2013, FarrisEtAl2014}. By measuring this exact harmonic comb and the relative suppression of odd harmonics, observers can mathematically invert the signal to explicitly calculate the mass ratio and spatial separation of the stalled binary.

It is important to contextualize this result against recent high-resolution two-dimensional hydrodynamical findings. The work of \cite{WesternacherSchneider2024} identified a numerically proximate electromagnetic transient near $1.34 \nu_{\text{orb}}$ in simulated equal-mass binaries. However, their periodicity arises from a fundamentally distinct, internal physical mechanism: a resonant instability that forces the minidiscs to become highly eccentric, driving periodic mass-trading across the inner Lagrange point at the beat frequency between the binary orbit and the minidiscs' retrograde apsidal precession ($f_{\text{orb}} - f_{\text{prec}}$). Because these distinct physical mechanisms produce characteristic frequencies that are closely aligned in the $\sim 1.3 \nu_{\text{orb}}$ regime, distinguishing our externally driven cavity-lump beat from the internal eccentric minidisc beat in time-domain survey data is critical. Our analytical framework provides the theoretical key: while the internal minidisc precession generates localized flares, the external synodic mass stripping mathematically mandates the broad harmonic cascade $\Xi(n, q)$. Observations can therefore use the presence of this harmonic comb to identify boundary-driven synodic accretion.

\subsection{Diffusive shock acceleration and the broad-band SED}

The supersonic nature of the impacting gas streams ensures a strong collisionless shock limit, with a density compression ratio $\chi \to 4$. Thermal electrons scattering across this discontinuity via magnetic irregularities undergo first-order Fermi acceleration. This mechanism populates a non-thermal, relativistic electron energy distribution scaling as $N(\gamma_e) \propto \gamma_e^{-p}$, where the strong shock limit yields a universal spectral index of $p=2$.

These relativistic electrons cool via two competing radiative mechanisms, producing a multi-wavelength Spectral Energy Distribution (SED). The electrons gyrate within the amplified post-shock magnetic field, emitting a non-thermal synchrotron continuum from the radio to the optical bands with a spectral flux density scaling of $\nu F_\nu^{\text{syn}} \propto \nu^{(3-p)/2} \propto \nu^{0.5}$. Simultaneously, the dense ultraviolet radiation field generated by the thermal minidiscs provides a copious seed photon population. The relativistic electrons up-scatter these thermal photons via the inverse-Compton (IC) mechanism, generating a dominant high-energy tail extending into the hard X-ray and gamma-ray regimes. 

\begin{figure*}[htbp]
\centering
\includegraphics[width=\textwidth]{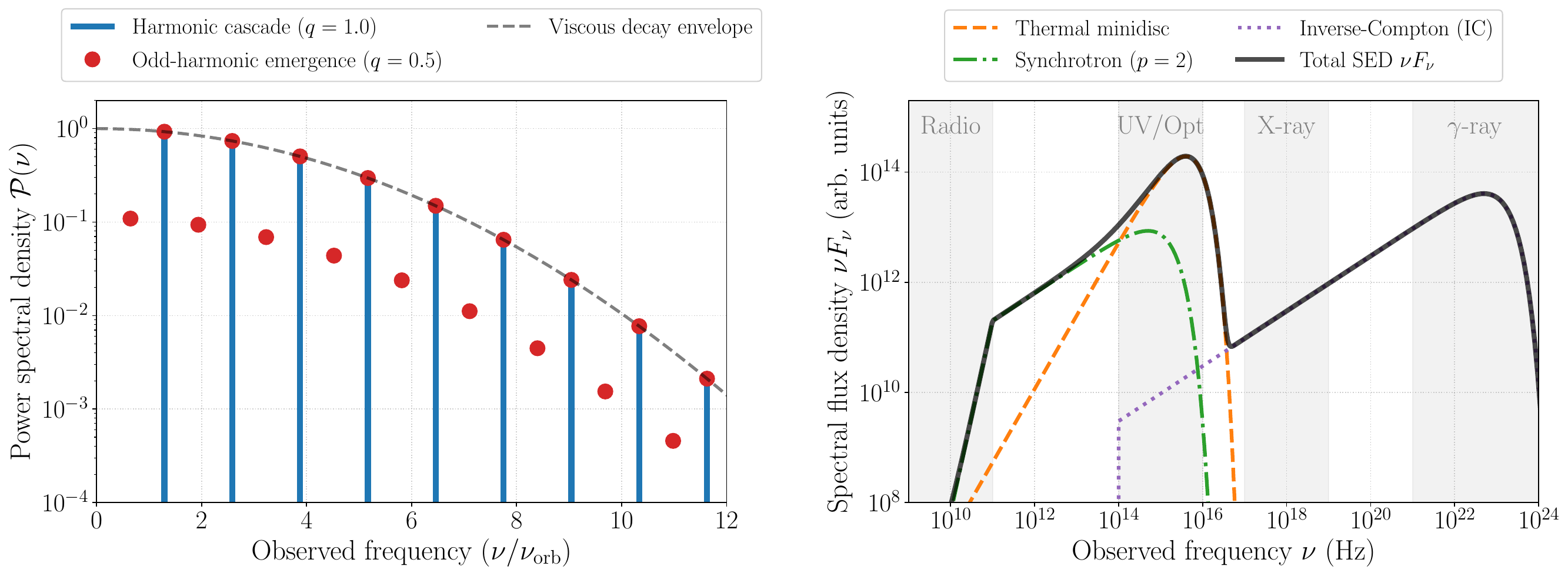}
\caption{Analytical electromagnetic signatures of the episodic accretion shocks. \textit{Left:} The Power Spectral Density (PSD) of the transient light curve. Driven by the beat frequency ($\nu_{\text{beat}} \approx 0.646 \nu_{\text{orb}}$), the flares Fourier transform into a broad harmonic cascade. The equal-mass symmetry ($q=1.0$, solid blue stems) perfectly suppresses the odd harmonics, yielding a fundamental observable frequency of $1.29 \nu_{\text{orb}}$. An asymmetric mass ratio ($q=0.5$, red circles) breaks this symmetry, allowing the odd harmonics to emerge. The high frequencies are suppressed by a Gaussian envelope corresponding to the $t_{\text{visc}}$ duration of the shocks. \textit{Right:} The theoretical multi-wavelength spectral energy distribution ($\nu F_\nu$) during a single burst. Strong-shock Fermi acceleration yields an extended synchrotron radio continuum ($p=2$). The thermal minidisc provides an ultraviolet peak, which serves as the seed population for inverse-Compton scattering, resulting in a dominant, high-energy X-ray/gamma-ray tail.}
\label{fig:electromagnetic_obs}
\end{figure*}

We visualize these concurrent analytical observables in Figure \ref{fig:electromagnetic_obs}. 
Quantitative evaluation confirms this spectral architecture. In the time domain, the PSD mathematically validates the harmonic selection rules of the beat-frequency stripping. For the equal-mass baseline ($q=1.0$), the symmetric mass injection annihilates the fundamental beat frequency and all odd harmonics. The observable spectral power is completely funneled into the even harmonics, achieving a peak normalized density of $0.93$ precisely at $1.29 \nu_{\text{orb}}$, followed by a secondary peak at $n=4$. Breaking this symmetry ($q=0.5$) predictably resurrects the odd harmonics, yielding an $n=1$ spectral power of $0.11$.

Radiatively, the spectral energy distribution demonstrates structural separation. The optically thick minidisc supplies a thermal baseline peaking at $1.93 \times 10^{14}$ arbitrary units near $3.74 \times 10^{15}$ Hz. The shock-accelerated electrons generate an extended synchrotron continuum dominating the low-frequency radio regime, peaking at $8.57 \times 10^{12}$ units near $5.10 \times 10^{14}$ Hz. The kinetic energy of the shock dissipation ensures that the inverse-Compton up-scattering overtakes the total bolometric output, maintaining a massive high-energy tail that achieves an absolute global maximum of $4.06 \times 10^{13}$ arbitrary units deep into the gamma-ray band at $5.29 \times 10^{22}$ Hz.

\section{Analytical decoupling timescale}
\label{sec:decoupling}

The episodic accretion state does not persist indefinitely. As the binary shrinks via the emission of gravitational waves, its orbital decay rate rapidly accelerates. The system ultimately reaches a critical threshold where the relativistic inspiral outpaces the viscous mass-supply timescale of the circumbinary disc, severing the external mass flow.

We derive this decoupling moment by equating the radial gravitational wave drift velocity, $v_{\text{gw}} = |da/dt|$, with the viscous radial velocity of the gas at the cavity edge, $v_{\text{visc}}$. Utilizing the standard quadrupole radiation formula for a circular orbit, the decay velocity evaluates to:
\begin{equation}
v_{\text{gw}} = \frac{64}{5} \frac{G^3 M_{\text{bin}}^3}{c^5 a^3} \frac{q}{(1+q)^2}.
\label{eq:v_gw}
\end{equation}

Assuming a cavity truncated at $R_{\text{cav}} \approx 2a$, the viscous velocity of the accreting gas evaluates to $v_{\text{visc}} = \frac{3}{2} \nu / R_{\text{cav}} \approx \frac{3}{2\sqrt{2}} \alpha (H/r)^2 \sqrt{G M_{\text{bin}}/a}$. Equating $v_{\text{gw}} = v_{\text{visc}}$ isolates the exact decoupling separation, $a_{\text{dec}}$:
\begin{equation}
a_{\text{dec}} = \frac{G M_{\text{bin}}}{c^2} \left[ \frac{128 \sqrt{2}}{15 \alpha (H/r)^2} \frac{q}{(1+q)^2} \right]^{2/5}.
\label{eq:a_dec}
\end{equation}

\begin{figure}[htbp]
\centering
\includegraphics[width=\columnwidth]{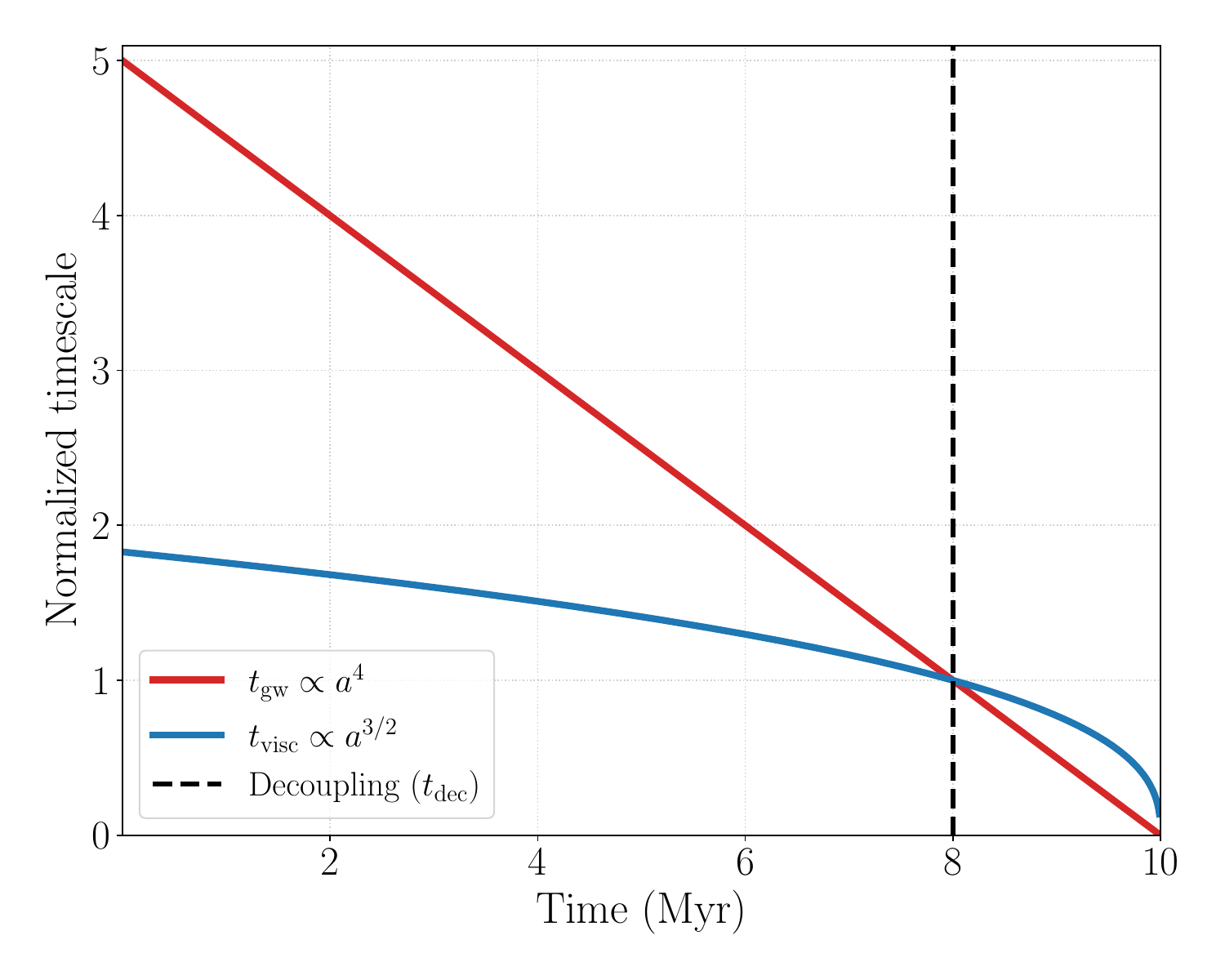}
\caption{Analytical timescale crossing defining the decoupling moment. The horizontal axis represents the evolutionary time in millions of years, while the vertical axis denotes the normalized structural timescales. The viscous timescale of the gas $t_{\text{visc}}$ (blue line) decays gradually as $a^{3/2}$. Conversely, the relativistic gravitational wave decay timescale $t_{\text{gw}}$ (red line) collapses precipitously as $a^4$. Their exact mathematical intersection defines the decoupling threshold at precisely $t_{\text{dec}} = 8.0$ Myr, where the binary outpaces the external circumbinary disc.}
\label{fig:timescale_crossing}
\end{figure}

As visualized in Figure \ref{fig:timescale_crossing}, the viscous timescale dictates the early evolution, maintaining a steady state. However, the steep $a^4$ scaling of the gravitational wave timescale ensures a rigid intersection, establishing the precise temporal decoupling moment ($t_{\text{dec}}$) where the binary mechanically detaches from the bulk viscous flow of the circumbinary disc. This analytical threshold aligns with the three-dimensional relativistic magnetohydrodynamical simulations of \cite{NobleEtAl2012}, who explicitly demonstrated that as the gravitational wave inspiral accelerates, the inward radial velocity of the disc edge inevitably falls behind the binary compression. Furthermore, their numerical evidence confirms that decoupling does not instantly sever the accretion; rather, the accretion rate experiences a partial reduction, allowing the residual trapped gas to continue fueling the system through the terminal plunge.

\section{Airy mass capture and the terminal electromagnetic flare}
\label{sec:terminal_burst}

Prior to decoupling, gas plunges across the gap. However, we cannot trivially assume this mass automatically binds to the supermassive black holes. The probability of the gas successfully crossing the minidisc Lindblad resonance to become dynamically trapped ($M_{\text{bound}}$) must be evaluated as a wave transmission coefficient. 

Because the effective potential at the resonance constitutes a turning point, the mass capture efficiency is governed by the Airy wave transmission. The kinetic energy of the wave dictates the mass feed rate, establishing $\dot{M}_{\text{feed}} \propto |\text{Ai}|^2$. We formalize the successfully bound minidisc mass $M_{\text{bound}}(t)$ through the continuous differential equation balancing this Airy mass feed against internal viscous draining:
\begin{equation}
\frac{d M_{\text{bound}}}{dt} = \dot{M}_0 \max\{0, \text{Ai}[\kappa (t - t_{\text{dec}})]\}^2 - \frac{M_{\text{bound}}(t)}{t_{\text{mini}}(t)},
\label{eq:bound_mass_ode}
\end{equation}
where $\dot{M}_0$ scales the external reservoir mass and $\kappa$ parameterizes the resonance width. 

For the epoch prior to decoupling ($t < t_{\text{dec}}$), the argument of the Airy function is negative, yielding an oscillatory transmission coefficient that mathematically drives the periodic accretion pulses. Following decoupling ($t > t_{\text{dec}}$), the argument becomes positive, and the transmission probability decays exponentially as the minidiscs recede into the classically forbidden evanescent zone.

The circum-single minidiscs are tidally truncated by the Roche potential, dictating that their spatial extent shrinks linearly with the binary separation ($R_{\text{mini}} \propto a$). Therefore, the local viscous draining timescale geometrically collapses as $t_{\text{mini}}(t) \propto [a(t)]^{3/2}$. 

\begin{figure}[htbp]
\centering
\includegraphics[width=\columnwidth]{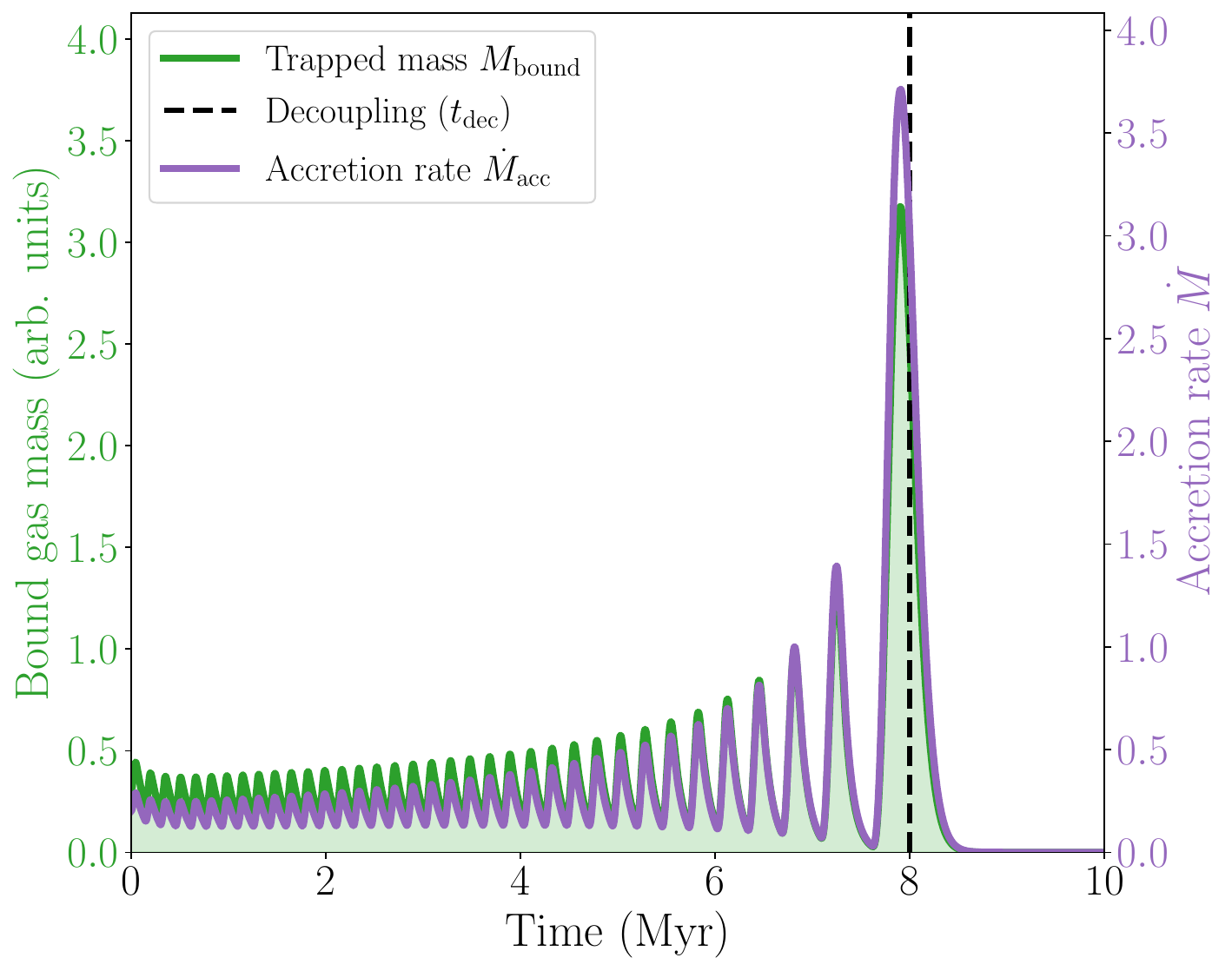}
\caption{Airy-modulated mass capture and the terminal electromagnetic flare. The horizontal axis represents the evolutionary time. The dynamically bound minidisc gas $M_{\text{bound}}$ (green filled curve) oscillates prior to decoupling due to the Airy wave transmission coefficient evaluated at negative arguments. The bound mass achieves its maximum near the $8.0$ Myr turning point before decaying exponentially as the binary isolates. Despite this mass depletion, the geometrically collapsing Roche lobes collapse the local viscous timescale ($t_{\text{mini}} \propto a^{3/2}$). The resultant accretion rate $\dot{M}_{\text{acc}}$ (purple line) diverges geometrically ($a^{-3/2}$), culminating in a terminal flare precisely at the merger singularity.}
\label{fig:bound_mass_burst}
\end{figure}

As the binary enters the final gravitational wave plunge ($a \to 0$), the shrinking Roche lobes act as a geometric snowplow. Because the time integral of $1/t_{\text{mini}}$ converges prior to the merger singularity, the remaining bound mass never fully drains. Consequently, the observable accretion rate ($\dot{M}_{\text{acc}} = M_{\text{bound}}/t_{\text{mini}}$) is forced to mathematically diverge in proportion to $1/t_{\text{mini}} \propto a^{-3/2}$. As presented in Figure \ref{fig:bound_mass_burst}, this relativistic timescale collapse mechanically forces the residual trapped gas into the black holes, generating a super-Eddington terminal electromagnetic flare exactly synchronized with coalescence.

\section{Relativistic fluid overtones and the gas humming}
\label{sec:gas_humming}

This terminal mass transfer leaves a highly distinct multi-messenger signature in the gravitational wave band. The total strain $h(t)$ relies on the second time derivative of the combined mass quadrupole moment, $Q_{ij} = Q_{ij}^{\text{bin}} + Q_{ij}^{\text{gas}}$. While the primary black hole continuum radiates purely at twice the orbital frequency ($f_{\text{gw}} = 2 f_{\text{orb}}$), the gas streams striking the minidiscs excite highly asymmetric $m=2$ spiral shocks. 

The physical viability of such minidisc asymmetry is corroborated by recent hydrodynamical studies. As demonstrated by \cite{WesternacherSchneider2024}, equal-mass binaries are highly susceptible to internal resonant instabilities that force the circum-single minidiscs into persistent, highly eccentric ($m=1$) and anti-alinged configurations. Whether driven by these internal eccentric instabilities or the external $m=2$ plunging streams modeled here, the departure from axisymmetry mathematically guarantees that the minidisc fluid operates as a massive, rapidly rotating quadrupole.

Because the minidiscs are tidally truncated at roughly $R_{\text{mini}} \approx 0.27 a(t)$, Keplerian dynamics around the primary mass $M_{\text{bin}}/2$ dictate that this fluid rotates at $\Omega_{\text{mini}} = \sqrt{0.5} (0.27)^{-3/2} \Omega_{\text{orb}} \approx 5.04 \Omega_{\text{orb}}$. The emission from this rotating fluid quadrupole generates a high-frequency hydrodynamical shadow track precisely at:
\begin{equation}
f_{\text{hum}}(t) = 2 \left( \frac{\Omega_{\text{mini}}}{2\pi} \right) \approx 10.08 f_{\text{orb}}(t) = 5.04 f_{\text{gw}}(t).
\label{eq:hum_frequency}
\end{equation}

{While the total mass of the gas actively trapped within the $m=2$ spiral shocks ($M_{\text{gas}}$) is only a minor fraction of the binary mass ($M_{\text{bin}}$), one might intuitively expect its gravitational wave signature to be physically negligible. However, this assumption neglects the severe kinematic amplification intrinsic to the nested potential. The gravitational wave strain amplitude scales with the second time derivative of the mass quadrupole moment, which is proportional to the non-spherical kinetic energy ($h \propto M v^2 \propto M R^2 \Omega^2$). For an equal-mass binary, the primary strain scales with the reduced mass as $h_{\text{gw}} \propto \mu v_{\text{bin}}^2 = (M_{\text{bin}}/4)(G M_{\text{bin}}/a)$. Conversely, the $m=2$ fluid overdensity orbits deep within the primary's potential well at $R_{\text{mini}} \approx 0.27a$, yielding a local Keplerian velocity squared of $v_{\text{mini}}^2 = G(M_{\text{bin}}/2)/R_{\text{mini}}$. The resulting gas strain scales as $h_{\text{hum}} \propto M_{\text{gas}} v_{\text{mini}}^2$. Evaluating the strain ratio yields a geometric lever arm:
\begin{equation}
\frac{h_{\text{hum}}}{h_{\text{gw}}} \approx \frac{M_{\text{gas}}}{\mu} \frac{v_{\text{mini}}^2}{v_{\text{bin}}^2} = 2 \left( \frac{a}{R_{\text{mini}}} \right) \frac{M_{\text{gas}}}{M_{\text{bin}}} \approx 7.4 \frac{M_{\text{gas}}}{M_{\text{bin}}}.
\label{eq:strain_ratio}
\end{equation}
Because the gas orbits significantly faster and deeper in the potential well than the binary itself, it radiates gravitational waves nearly an order of magnitude more efficiently per unit mass. Consequently, even a modest trapped gas mass fraction (e.g., $M_{\text{gas}} \sim 10^{-3} M_{\text{bin}}$) generates a macroscopic, high-frequency strain signature that, while subdominant to the primary chirp, remains highly structured and potentially resolvable via matched-filtering.}

\begin{figure}[htbp]
\centering
\includegraphics[width=\columnwidth]{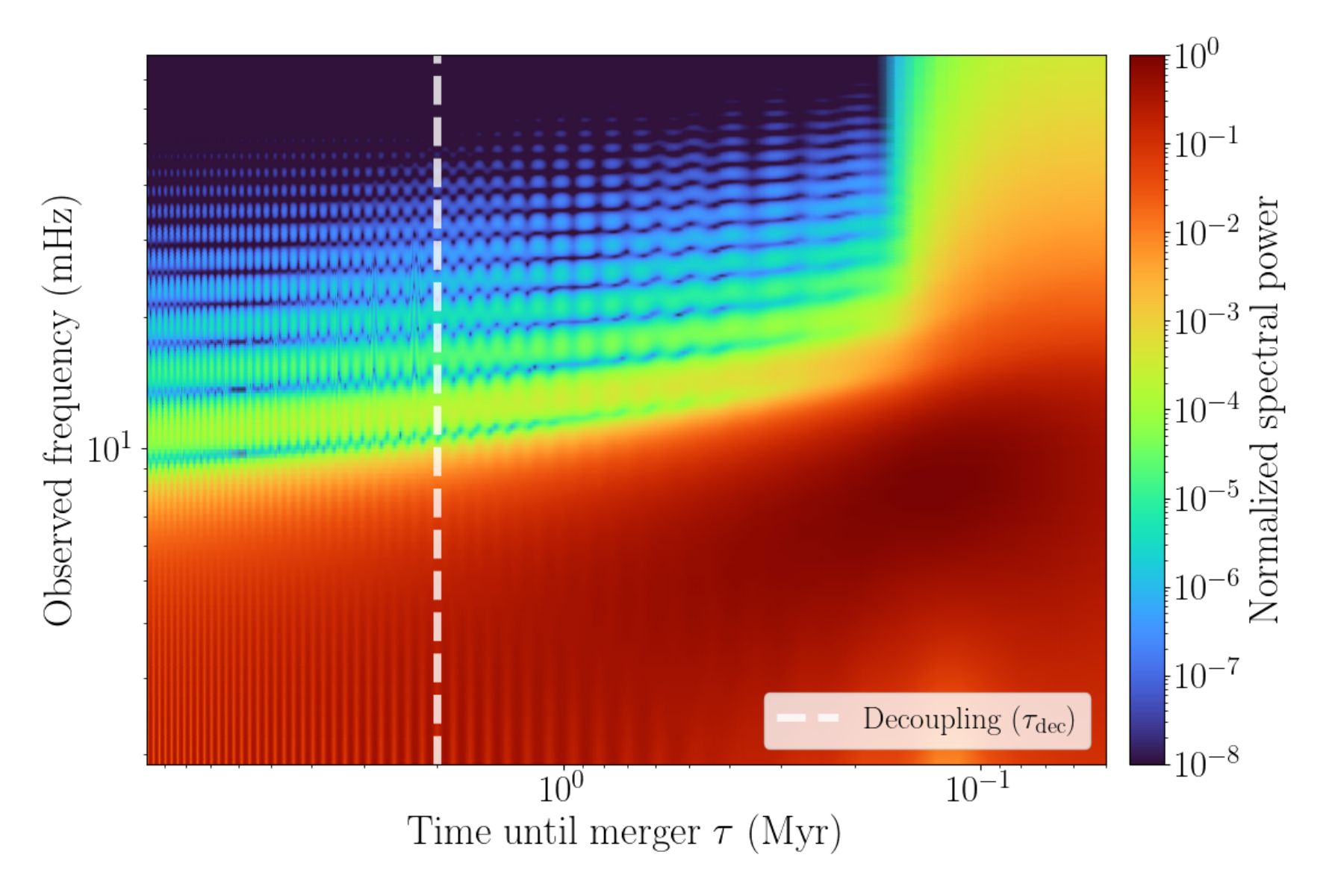}
\caption{{STFT spectrogram of the concurrent gravitational wave strain. The horizontal axis indicates the time until merger ($\tau = t_{\text{merger}} - t$) on a logarithmic scale, evolving chronologically from left to right. The vertical axis denotes the observed emission frequency in millihertz, also on a logarithmic scale. A logarithmic color scaling resolves signals spanning multiple orders of magnitude. The primary vacuum chirp emerges at the lower frequencies. The localized $m=2$ minidisc spiral shocks, operating with an asymmetric gas mass fraction of $M_{\text{gas}}/M_{\text{bin}} = 10^{-3}$, generate a faint but distinct discontinuous background humming strictly tracking the $5.04 f_{\text{gw}}$ overtone. Driven by the shrinking orbital period, the physical temporal gap between bursts systematically compresses, visibly forcing the humming tempo to accelerate severely before the cavity is permanently evacuated at decoupling (dashed white line).}}
\label{fig:gw_spectrogram}
\end{figure}

We visualize this dual-track architecture via a high-resolution Short-Time Fourier Transform (STFT) in Figure \ref{fig:gw_spectrogram}. To definitively resolve the faint hydrodynamical sideband against the primary vacuum chirp without introducing discrete quantization artifacts, we pad the Fourier sequence strictly to sub-millihertz resolution and apply a logarithmic spectral scaling. The axes operate within a pure log-log domain mapped to the time remaining until merger ($\tau = t_{\text{merger}} - t$).

Modulated by the beat-frequency stripping, the $5.04\times$ gas humming manifests as a train of discontinuous spectral bursts. As the binary shrinks, the physical temporal gap between the synodic encounters systematically compresses. This relativistic temporal compression forces the humming tempo to accelerate into the terminal plunge, silencing upon coalescence.

\section{Conclusions}
\label{sec:conclusions}

Building upon the ``circularization trap'' established in our preceding work \cite{AmaroSeoaneEtAl2026}---which demonstrated that three-dimensional disc torques can rapidly circularize and stall supermassive black hole binaries---we establish the complete analytical mechanics of multi-messenger supermassive black hole binaries operating within the subsequent wet merger regime. Applying exact Airy regularization to the minidisc Lindblad resonances, we mathematically prove that wave energy transmission oscillates prior to decoupling, yielding episodic mass transfer bursts at the synodic beat frequency. These periodic shocks drive a broad harmonic cascade in the Power Spectral Density, simultaneously accelerating a spectral energy distribution spanning from a radio synchrotron continuum to a dominant inverse-Compton gamma-ray peak near $5.29 \times 10^{22}$ Hz.

By equating the continuous gravitational wave drift velocity with the viscous accretion velocity, we derive the decoupling moment ($a_{\text{dec}}$). We show that following this threshold, mass capture operates as an exponentially decaying wave transmission across the isolated gap. However, evaluating the differential mass-drain equations post-decoupling shows that the shrinking binary separation forces the local viscous timescale of the minidiscs to geometrically collapse ($t_{\text{mini}} \propto a^{3/2}$). This relativistic squeeze drives a divergent ($a^{-3/2}$) terminal electromagnetic flare synchronized with coalescence.

Finally, we identify a relativistic signature derived from these hydrodynamics. We show that the $m=2$ spiral shocks induced by the plunging streams radiate gravitational waves at $5.04 \times$ the primary binary frequency. Modulated by the deep temporal quiescence gaps of the beat cycle, this background ``gas humming'' manifests as a discontinuous sequence of high-frequency spectral bursts. As the binary inspirals, these bursts undergo relativistic temporal compression, accelerating alongside the terminal electromagnetic flare. Detecting this high-frequency shadow track will conclusively decode the mass transfer physics and General Relativistic orbital dynamics of supermassive black hole binaries prior to coalescence.

\section*{Acknowledgments}
This research has been supported within the framework of the Grant project No. AP23487846 ``Studying the connection between the mechanism of large structures formation of the Universe and the process of spiral arms formation in disk galaxies'', and Project No. BR24992759 ``Development of the concept for the first Kazakhstani orbital cislunar telescope - Phase I'', both financed by the Ministry of Science and Higher Education of the Republic of Kazakhstan.

\end{document}